\documentclass[%
 reprint,
 amsmath,amssymb,
 aps,prl,
floatfix
]{revtex4-1}

\usepackage{graphicx}
\usepackage{dcolumn}
\usepackage{bm}

\usepackage[utf8]{inputenc}
\usepackage[english]{babel}
\usepackage[T1]{fontenc}
\usepackage{amssymb,amsmath,amsthm}
\usepackage{color,xcolor}
\usepackage{siunitx}
\usepackage{xstring}
\usepackage{newtxtext}

\usepackage{lmodern,pdftexcmds}
\usepackage[normalem]{ulem}
\definecolor{darkblue}{rgb}{0,0,.5}
\usepackage[linktocpage, colorlinks=true, linkcolor=darkblue, citecolor=darkblue]{hyperref}

\newcommand{\de}{\partial}

\newcommand{\vc}[1]{\bm{#1}}
\newcommand{\vt}[1]{\bm{\mathsf{#1}}}\DeclareMathAlphabet{\mathbfsf}{\encodingdefault}{\sfdefault}{bx}{sl}

\newcommand{\tsp}{\mathsf{T}}

\newcommand{\tr}{\operatorname{\mathrm{tr}}}

\newcommand{\dvg}{\operatorname{\mathrm{div}}}

\newcommand{\cauchy}{\vt{T}}

\newcommand{\extra}{\vt{S}}
\newcommand{\LV}{\vt{L}}

\newcommand{\F}{{\vt{F}}}

\newcommand{\edot}{\dot{\varepsilon}}

\newcommand{\Rey}{\mathrm{Re}}

\begin{document}

\title[Shear jamming and fragility of suspensions in a continuum model with elastic constraints]{Shear jamming and fragility of suspensions in a continuum model with elastic constraints}

\author{Giulio G. \surname{Giusteri}}
\email{giulio.giusteri@math.unipd.it}
\affiliation{Dipartimento di Matematica, Universit\`a degli Studi di Padova, Via Trieste 63, 35121, Padova, Italy}
\author{Ryohei \surname{Seto}}
\email{seto@wiucas.ac.cn}
\affiliation{Wenzhou Institute, University of Chinese Academy of Sciences, Wenzhou, Zhejiang 325001, China}
\affiliation{Oujiang Laboratory (Zhejiang Lab for Regenerative Medicine, Vision and Brain Health), Wenzhou, Zhejiang 325001, China}
\affiliation{The Graduate School of Information Science, University of Hyogo, Kobe, Hyogo 650-0047, Japan}

\date{\today}

\begin{abstract}
Under an applied traction, highly concentrated suspensions of solid particles in fluids can turn from a state in which they flow to a state in which they counteract the traction as an elastic solid: a shear-jammed state.
Remarkably, the suspension can turn back to the flowing state simply by inverting the traction.
A tensorial model is presented and tested in paradigmatic cases.
{We show that, to reproduce the phenomenology of shear jamming in generic geometries, it is necessary} to link this effect to the elastic response supported by the suspension microstructure rather than to a divergence of the viscosity.
\end{abstract}

\maketitle

The rheology of highly concentrated suspensions of solid particles dispersed in a viscous fluid features a number of surprising phenomena \citep{Guazzelli_2018,Morris_2020,Denn_2014} among which shear jamming raises important questions for its interpretation and challenges for its mathematical modeling.
If the concentration of particles is not very high, 
the suspension presents a fluid-like behavior.
At high concentration, one can instead observe a sudden solidification that occurs after some strain in a shear deformation, whence the name of shear jamming.

Under those conditions, the constant stress applied to the suspension is balanced by the elastic response of the solidified medium, 
arguably sustained by the network of contacts developed 
among the solid particles during the initial flow~\citep{Cates_1998a,Bi_2011,Vinutha_2016,Malkin_2020}.
The shear-jammed material is in a fragile state: if the applied stress is removed no motion arises, but if we reverse the stress, pushing in a sufficiently different direction, the suspension flows again and stops only once a certain strain is accumulated.
This history-dependent or protocol-dependent rheological response marks a key difference with states of isotropic jamming, in which the suspension is so concentrated to be unable to flow, irrespective of the direction of the applied forces.

We present a constitutive model that is able to reproduce the phenomenology of shear jamming and fragility. 
We base its construction on the understanding that shear jamming corresponds to the onset of an elastic response, 
related to geometric constraints at the micro-scale, 
rather than to a boost in dissipative phenomena, as implied by models containing a divergence of the viscosity.
We devised an effective and yet mathematically simple model, that features a small number of parameters easily linked to experimental measurements.

We define a tensorial model at the outset, without going through the process of designing a scalar model---typically tailored to a restricted set of motions---and then extending it.
In this way, our model is readily applicable to flows in general two- and three-dimensional geometries, both boundary- and pressure-driven ones.

In the last decade, a few models to describe the physics of dense suspensions have been proposed, with a focus on capturing discontinuous shear thickening and rationalizing the role of frictional contacts between the solid particles~\citep{Fernandez_2013, Seto_2013a, Heussinger_2013, Lin_2015, Ness_2016a, Clavaud_2017, Singh_2020}.
A common feature of such models is the attempt at building a direct link between the emergent behavior and the microstructure of the fluid, as characterized by the analysis of data coming form detailed simulations at the micro-scale level~\citep{Mari_2014,Seto_2019a}.
In relation to shear jamming, these models are not readily applicable to the macro-scale simulation of this phenomenon, since they identify the jammed state with a divergence of the viscosity \citep{Wyart_2014}
rather than the onset of elastic responses, so that mathematical singularities appear in the equations.
On the other hand, we can find important results in the construction of continuum models that focus on the macro-scale dynamics~\citep{Baumgarten_2019} and utilize a two-material approach, by following the coupled evolution of homogenized fluid and solid phases.
Such models are quite effective in reproducing certain observations, but feature several parameters that require calibration and involve a set of equations with considerable complexity.


Our model is intended to capture the suspension behavior at a rather large scale, 
when a single-fluid model is appropriate.
We do not take explicitly into account particle migration phenomena~\citep{Nott_1994,Morris_1999,Chacko_2018a} and rate-dependence of the viscosity or stress-induced solidification~\citep{Nakanishi_2012,Wyart_2014,Gillissen_2019}.
These can be incorporated for specific applications as extensions of the model but are not essential to reproduce shear jamming.
In fact, we stress that it is possible to give a very good qualitative description of shear jamming assuming a constant viscosity, by associating jamming with the appearance of elasticity.


The phenomenon of shear jamming can be viewed as the emergence of solidity due to the evolution of the suspension microstructure.
The activation of frictional contacts between the particles leads to the presence of percolating stable formations that span macroscopic portions of the system~\citep{Henkes_2010,Bi_2011,Vinutha_2016,Sarkar_2016,Zhao_2019}.
This microscopic non-locality of the internal interactions marks the transition from a regime in which momentum is transferred slowly and diffusively (viscous fluid) to a regime in which momentum travels fast and elastically across the system (jammed solid).
When the elastic response is very stiff, one can even approach the macroscopic non-locality represented by rigid-body motions.


Another important aspect brought at the forefront by shear jamming is the material memory.
We obviously observe a long-lasting memory of what would be the relaxed configuration in the jammed solid regime, 
but there is also a memory in the microstructure evolution that governs the type and amount 
of deformation possible within the fluid regime, 
in which no persistent elasticity is detected~\citep{GadalaMaria_1980,Seto_2019a}.
Both these aspects need to be captured in an effective continuum theory and 
we propose to use tensorial models for all of them.
As we shall see,
the kinematic descriptors of the system that are useful for our purposes are the velocity field and its gradient, 
to capture the viscous dissipation, and a tensorial measure of the strain induced on the material by the motion.
The latter quantity keeps track of the microstructural deformation and, by limiting its evolution with unilateral constraints in the appropriate space of tensors, we can capture the transition between the fluid regime and the solid one, meanwhile preserving the characteristic reversibility represented by the fragility of the shear-jammed state.



\medskip

\textit{The tensorial model.}---%
A crucial role in shear jamming is played by the history of the deformation, since it induces some organization of the suspension microstructure, eventually responsible for the solid-like behavior.
Alongside the evolution equation for the velocity field $\vc u$ of a fluid with mass density $\rho$,
\begin{equation}
\rho\bigg(\frac{\de\vc u}{\de t}+(\vc u\cdot\nabla)\vc u\bigg) =\dvg\cauchy,
\end{equation}
driven by the Cauchy stress tensor $\cauchy$, we consider the evolution equation for the deformation gradient tensor $\F$ \citep[see][Chap.~3.2]{Phan-Thien_2017} in \emph{spatial coordinates}:
\begin{equation}\label{eq:F}
\frac{\de\F}{\de t}+(\vc u\cdot\nabla)\F=(\nabla\vc u)\F.
\end{equation}
Equation \eqref{eq:F} is an exact kinematic relation between the velocity and the displacement of fluid elements and does not contain any constitutive assumption.

From $\F$ we define $\vt B \equiv \F\F^\tsp$ and $\LV \equiv \tfrac{1}{2}\log\vt B$,
where $\log$ denotes the matrix logarithm. This is well-defined because the left Cauchy--Green tensor $\vt B$ is symmetric and positive definite for any physical motion.
These kinematic quantities track the local strain by factoring out rigid rotations, that should not affect the elastic response.
The tensor $\LV$ is the spatial counterpart of the Hencky strain and a generalization of the scalar strain measured in simple shear flows.
{Several advantages of its use are discussed in Ref.~\cite{Neff_2016}.}

An important feature of $\LV$ is that it is traceless whenever $\det{\vt B}=1$. This is always the case for us, because we assume incompressibility of the material, namely $\dvg\vc u=0$ at all times. We write the stress tensor as a pressure term plus the traceless extra stress $\extra$, so that $\cauchy = -p\vt I+\extra$.
The extra stress is the sum of a viscous dissipation plus an elastic response.
The dissipative term takes the  form $2 \eta \vt D$, wherein the effective viscosity $\eta$ of the suspension multiplies the symmetric part of the velocity gradient $\vt D \equiv (\nabla\vc u+\nabla\vc u^\tsp)/2$.


Regarding the elastic contribution to the stress, we assume that there exists a predetermined {subset} $\mathcal{N}$ 
in the space $\mathcal{S}$ of local strains (symmetric and traceless tensors) {corresponding to states in which} the material is \emph{elastically neutral}.
It means that, at each point $\vc x$ and instant $t$, if $\LV(\vc x,t)$ is in $\mathcal{N}$ there is no elastic response.
{This assumption is motivated by the observation that there is a regime in which particle contacts contribute to the effective viscosity but do not store elastic energy and the macroscopic response is purely viscous.
In our model}, the elastic response will be proportional to a suitable measure of how far $\LV(\vc x,t)$ is from $\mathcal{N}$.
The overall isotropy of the suspension suggests to take $\mathcal{N}$ to be a ball centered at the null tensor, namely
$\mathcal{N} \equiv \big\{\,\vt M\in\mathcal S: \Vert\vt M\Vert\leq r\,\big\}$
where, {for any arbitrary tensor $\vt M$, we set  $\Vert\vt M\Vert^2 \equiv {\tr(\vt M^\tsp\vt M)/2}$ and $r>0$ is a dimensionless material parameter that indicates how much the suspension needs to be sheared to achieve a jammed microstructure (and identifies the radius of $\mathcal N$).}
The value of $r$ would {typically be a decreasing function of} the volume fraction of solid particles.
{When $r=0$ the suspension is  an elastic solid, as under isotropic jamming.}

Since $\mathcal{N}$ is a closed convex subset of $\mathcal{S}$, a projection operator $\Pi:\mathcal{S}\to\mathcal{N}$ is well-defined and, for any $\vt M\in\mathcal S$, the tensor $\Pi(\vt M)$ is the element of $\mathcal N$ closest to $\vt M$. 
Such a projection can be easily expressed as
\begin{equation}
\Pi(\vt M) \equiv 
\begin{cases}
\vt M                   &\text{if }\Vert\vt M\Vert\leq r,\\
r\vt M/\Vert\vt M\Vert  &\text{if }\Vert\vt M\Vert> r.
\end{cases}
\end{equation}

To reflect the fact that an elastic response is activated whenever the logarithmic measure of strain $\LV$ leaves the neutral {subset} $\mathcal N$, we assume an extra stress of the form
\begin{equation}
\extra=2\eta\vt D+2\kappa\big(\LV-\Pi(\LV)\big),
\end{equation}
where the material parameter $\kappa>0$ represents an elastic stiffness.
This is a ``soft'' way of constraining the strain (as opposed to keeping it always within $\mathcal N$) that is able to better reproduce some details of the elastic effects observed in the proximity of jamming~\cite{Malkin_2020}.
With the present model, the strain of the jammed material tends to remain close to the boundary of $\mathcal N$ if the applied stress is driving it outwards. Conversely, the suspension can flow again as soon as the stress drives $\LV$ towards the interior of $\mathcal N$. In this way we can capture both shear jamming and the fragility of the jammed state.

{The tensor $\Pi(\LV)$ corresponds to a conformation tensor. It describes a microstructure that closely follows the strain $\LV$ up to the boundary of $\mathcal N$, where shear jamming prevents further microstructural deformations.}
The inclusion of additional dissipative phenomena, that may appear at the onset of jamming, can be achieved by letting $\eta$ depend on a parameter like $\lambda \equiv \Vert\LV-\Pi(\LV)\Vert$.


\medskip

\textit{Planar extensional flows.}---%
We highlight the basic features of the model in an idealized case, for which analytical computations can be carried out.
Under the deformation associated with planar extensional flows, the current position of a particle that occupies the place $(x_0, y_0)$ at time $0$ is given by 
$\vc\varphi(x_0, y_0,t)=(x_0e^{\varepsilon(t)},y_0e^{-\varepsilon(t)})$
and its spatial inverse is 
$\vc\varphi^{-1}( x, y,t)=(xe^{-\varepsilon(t)},ye^{\varepsilon(t)})$, where $\varepsilon(t)$ is an arbitrary function of time and measures the strain of the material.
We immediately obtain $\F=\mathrm{diag}(e^{\varepsilon(t)},e^{-\varepsilon(t)})$ and consequently $\vt B=\mathrm{diag}(e^{2\varepsilon(t)},e^{-2\varepsilon(t)})$.
In this case, the computation of the matrix logarithm is straightforward and yields $\LV=\mathrm{diag}(\varepsilon(t),-\varepsilon(t))$.

The velocity is $\vc u(t)=(\edot(t) x,-\edot(t) y)$ and the symmetric part of the velocity gradient, the usual measure of the rate of deformation, is
\begin{equation}
\vt D=\begin{pmatrix}
\dot{\varepsilon}(t)  & 0 \\
0  & -\dot{\varepsilon}(t)  \\
\end{pmatrix}=\frac{\de\LV}{\de t}.
\label{eq:Dext}
\end{equation}
We stress that the second identity in \eqref{eq:Dext} \emph{is not} valid for a generic flow 
(it does not hold, e.g., in simple shear);
when vorticity is present, rotation affects the deformation history in a nontrivial way, 
and $\vt D$ and $\LV$ cannot remain aligned.
This fact corresponds to the well known presence of normal stress differences in simple shear flows of viscoelastic fluids.

\begin{figure}[thb]
\centering
\includegraphics[width=86mm]{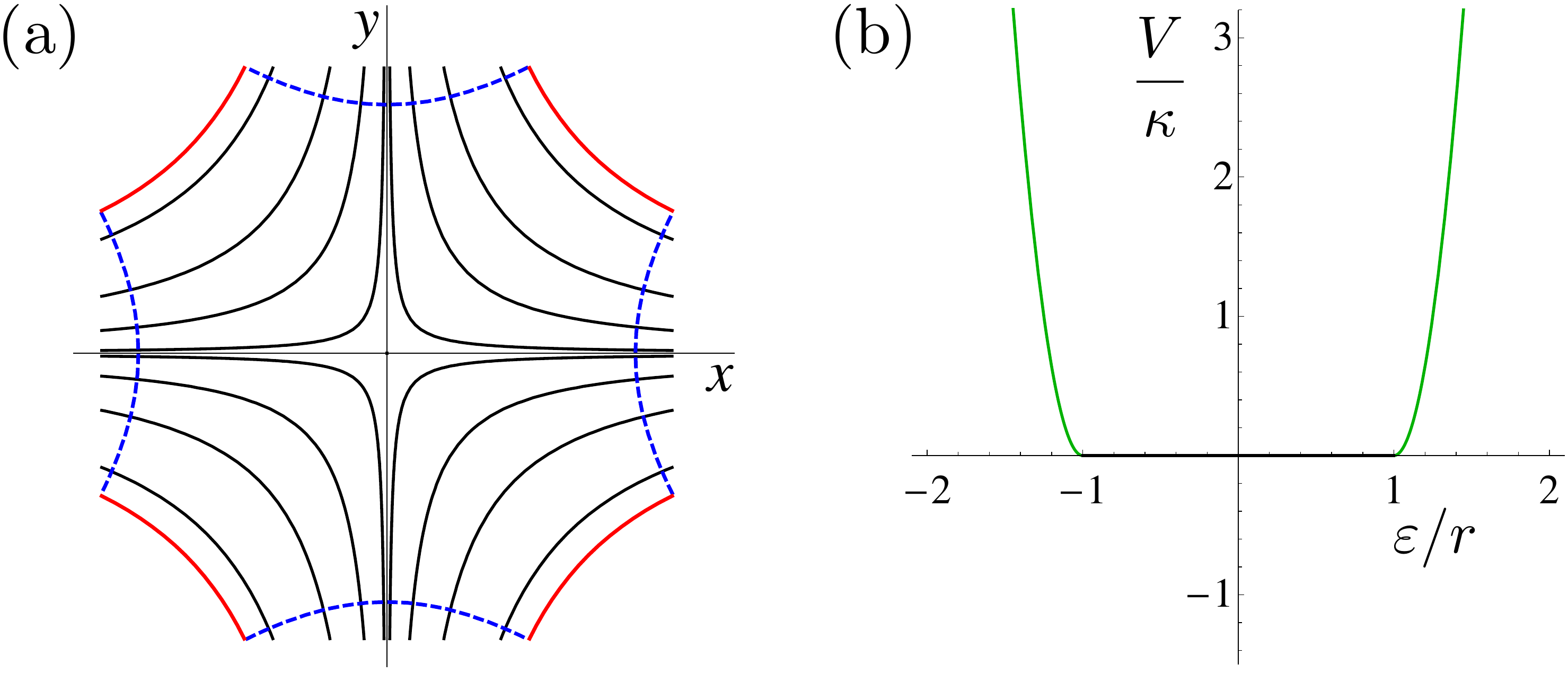}
\caption{We can imagine planar extension in a cross channel (a), with hyperbolic boundaries (red solid lines) that allow for perfect slip, to which we apply a pressure difference between top/bottom inlet and right/left outlet (blue dashed lines). The linearized equations for the proposed model reduce in this case to the scalar ordinary differential equation \eqref{eq:1d} for the strain function $\varepsilon$. {Considering the elastic force in Eq.~\eqref{eq:1d}, we see that it corresponds to that of a damped oscillator with (b) an elastic potential energy $V$ featuring a flat region for $\varepsilon \in [-r,r]$.}
}\label{fig:extensional}
\end{figure}

We consider the extensional flow in a cross channel with hyperbolic boundaries that allow for a perfect slip of the fluid (Fig.~\ref{fig:extensional}a).
A pressure difference applied to inlets and outlets of the channel generates normal tractions $\tau\vc n$ at outlets and $-\tau\vc n$ at inlets, where $\vc n$ is the unit outer normal to the boundary.
In a slow-velocity regime, the linearized flow equations give the pressure field $p(x,y,t)=\rho\ddot{\varepsilon}(t)(y^2-x^2)/2$.
The balance of stress at $(x,y)=(l,0)$ yields the following equation:
\begin{equation}\label{eq:1d}
\frac{\rho l^2}{2}\ddot{\varepsilon}+2\eta\dot{\varepsilon}-\tau
=
\begin{cases}
  -2\kappa [\varepsilon+r] & \text{if $\varepsilon <-r$},\\
                       +0  & \text{if $-r \leq \varepsilon \leq r$}, \\
  -2\kappa [\varepsilon-r] & \text{if $\varepsilon >r$},
\end{cases}
\end{equation}
{where the dimensionless parameter $r$ denotes, as above, the radius of the neutral subset $\mathcal N$.
The derivation of Eq.~\eqref{eq:1d} is reported in the Supplemental Material \cite{Note1}, Sec.~A.}

This is a scalar ordinary differential equation for the strain $\varepsilon(t)$, 
equivalent to that of a damped oscillator with elastic potential energy that  features a flat region for $\varepsilon \in [-r,r]$
and parabolic branches outside that interval (Fig.~\ref{fig:extensional}b). 
This entails transitions between a viscous fluid behavior, for $\varepsilon \in [-r,r]$ {when there is no elastic force}, and that of a viscoelastic solid {when elastic forces are activated}.


\begin{figure}[thbp]
\centering
\includegraphics[width=86mm]{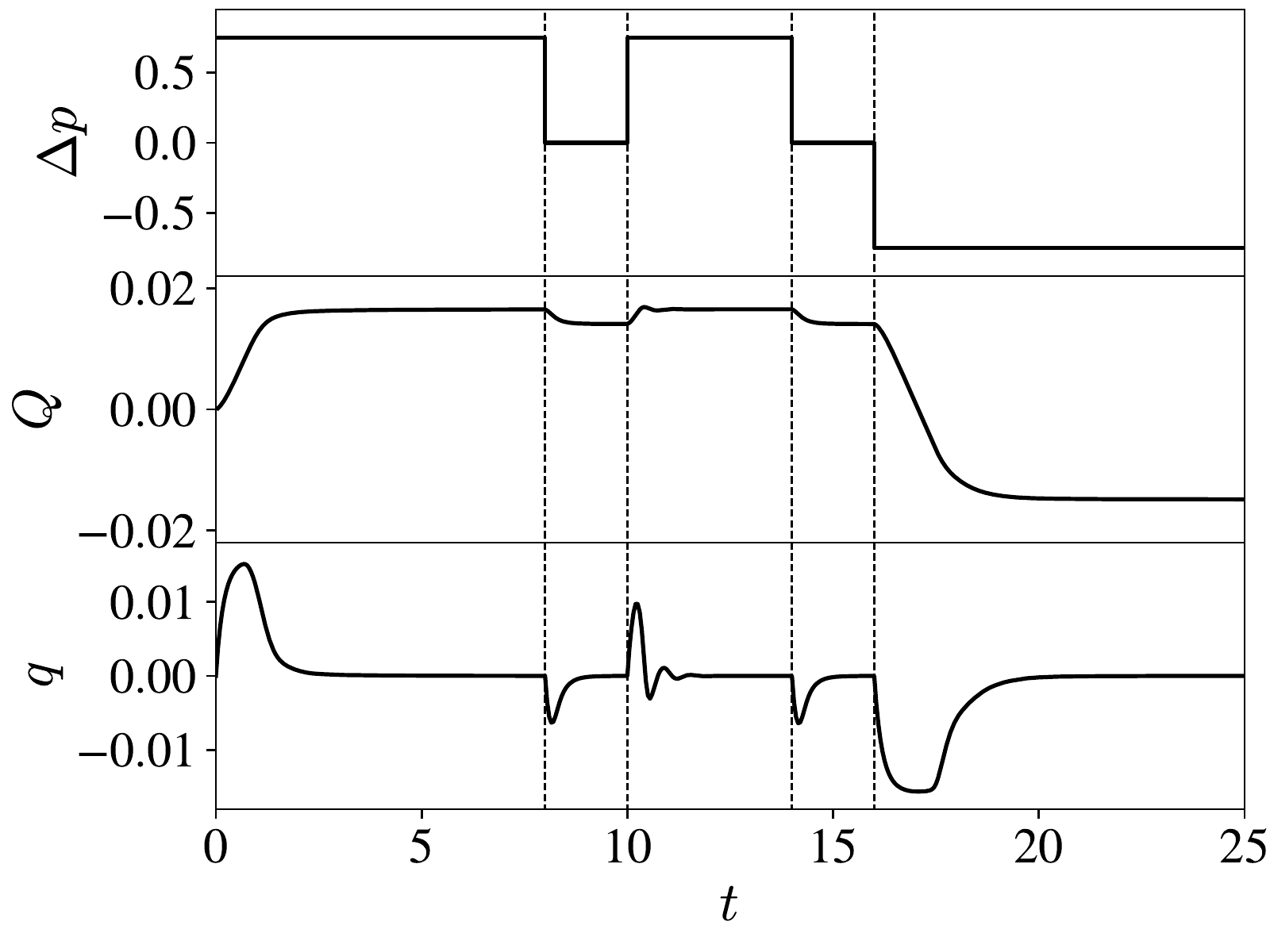}
\caption{%
The present model can reproduce the features of shear jamming in a complex flow
through a contraction.
We varied the pressure difference $\Delta p(t)$ imposed between the left and the right opening 
and measured the flow rate $q(t)$ through a cross section of the channel and its time integral $Q(t)=\int_0^t q(s)ds$.
Parameters in the simulation: {channel width equals the contraction length $\ell$, while the contraction width is $\ell/4$;} $\Rey=20\sqrt{0.75}\approx 17$, $\tilde\kappa=100/\sqrt{0.75}\approx 116$, and $r=1.5/\sqrt{2}$.
}\label{fig:ux}
\end{figure}

\begin{figure}[thbp]
\centering

\includegraphics[width=86mm]{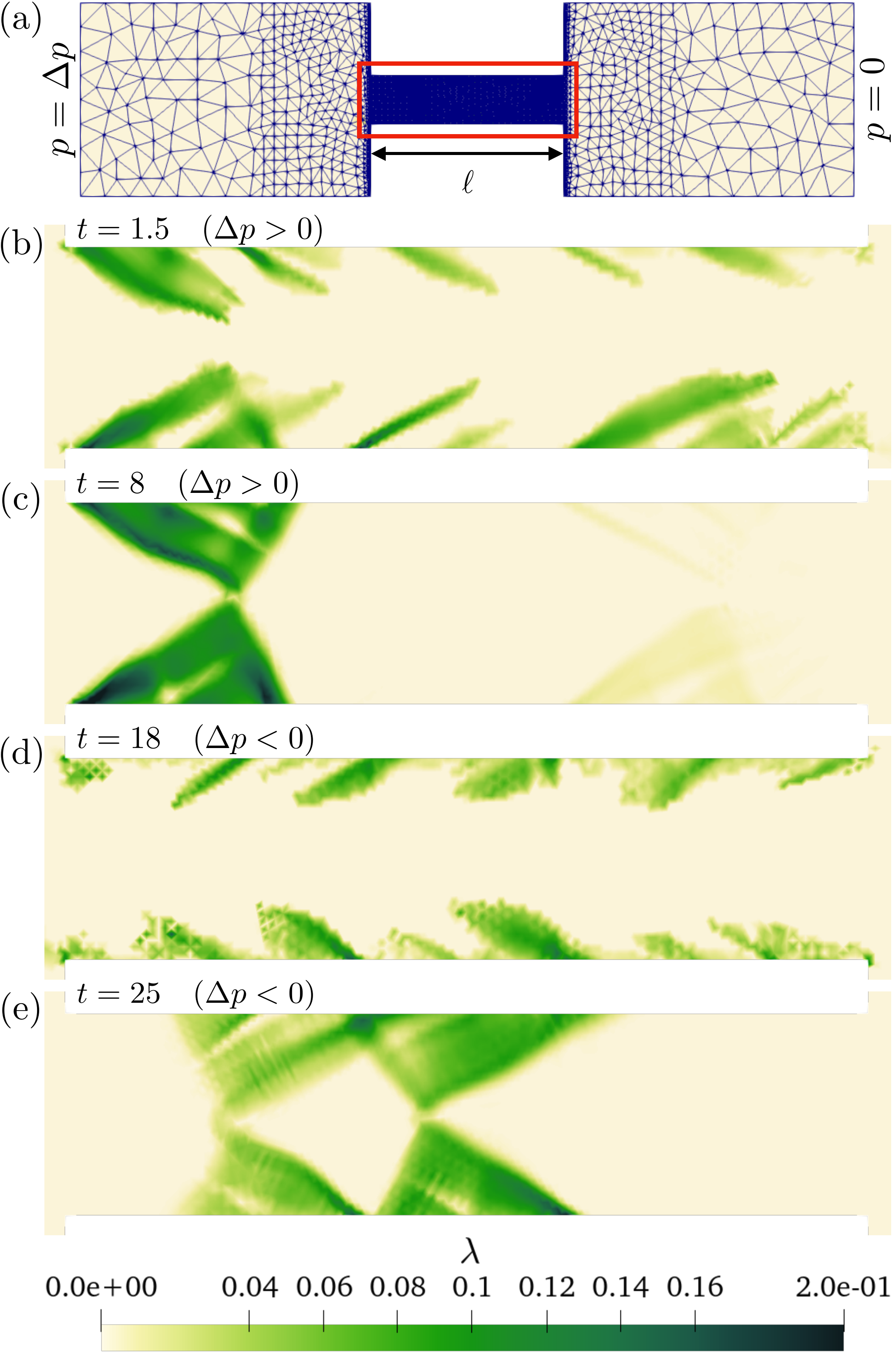}

\caption{(a) The entire domain is shown with the discretized mesh.
(b--e) The clogging of the channel is due to the presence, within the contraction to which images are restricted, of shear-jammed domains. These are characterized by a non-vanishing elastic response measured by the parameter 
$\lambda \equiv \Vert\LV-\Pi(\LV)\Vert$.
{From panels (b) and (d) we can see that the jammed domains nucleate and grow from the contraction boundaries, where the strain grows faster.}
There is a clear difference between the shear-jammed state (c) achieved at $t=\SI{8}{}$ with a positive $\Delta p$, that pushes rightward, and (e) the one obtained at $t=\SI{25}{}$ after reorganization with a negative $\Delta p$, that pushes leftward.
{In particular, the sawtooth shape of the jammed regions is reflected.}}
\label{fig:contraction}
\end{figure}

\medskip
\textit{Clogging and unclogging.}---%
Let us now consider how the model performs in simulating a paradigmatic pressure-driven flow through a contraction.
In this planar flow, 
{the maximum width of the channel (Fig.~\ref{fig:contraction}) equals the contraction length $\ell$, while the contraction width is $\ell/4$. The total length of the domain is $4\ell$ and we assume a uniform unit depth.}
Periodic boundary conditions for $\vc u$ and $\F$ are {imposed at the left and right boundary of the domain, while no-slip conditions are assumed on the top and bottom walls.
The pressure is not periodic: a pressure difference $\Delta p$ between the right and left openings is driving the flow.}

We introduce a dimensionless form of the evolution equations by defining a reference pressure difference $P$ and taking the channel width $\ell$ as reference length.
A reference time scale is $t_0 \equiv \ell\sqrt{\rho/P}$, leading to $\tilde t=t/t_0$.
From these, we set the Reynolds number $\Rey$ and {the dimensionless elasticity constant $\tilde\kappa$ according to}
\[
\Rey \equiv \frac{\ell\sqrt{\rho P}}{\eta}\qquad\text{and}
\qquad {\tilde\kappa \equiv \frac{2\kappa \ell}{\eta}\sqrt{\frac{\rho}{P}}},
\]
{and so that the dimensionless flow equation reads}
\begin{equation}
{\Rey\bigg(\frac{\de\tilde{\vc u}}{\de \tilde{t}}+(\tilde{\vc u}\cdot\nabla)\tilde{\vc u}\bigg) = -\nabla\tilde{p}+\nabla^2\tilde{\vc u}+\tilde\kappa\dvg\big(\LV-\Pi(\LV)\big).}
\end{equation}
In what follows we consider all quantities as dimensionless but 
drop the tildes for simplicity.

At startup, $\Delta p$ is set positive, the flow accelerates and the flow rate reaches a maximum at about $t = 1$ (Fig.~\ref{fig:ux}).
The deformation induced by the flow gives rise to shear-jammed domains that grow from the boundaries towards the center of the contraction {(Fig.~\ref{fig:contraction}b)}.
This activates an elastic response within the material that hinders the flow.
{The pressure drop in the clogged state ($t=8$) is sustained by jammed regions with a characteristic sawtooth shape (Fig.~\ref{fig:contraction}c). The intensity of the elastic response in each region depends on the local pressure drop.}
When we remove the pressure difference, from $t=8$ to $t=10$ (and again from $t=14$ to $t=16$), the stored elastic energy is completely released with a small recoil and then the flow stops (Fig.~\ref{fig:ux}). 
Nevertheless, the microstructure remembers to be close to shear jamming and when the pressure difference is turned on again ($t=10$) only a small fluid displacement is produced, since we assist to a rapid reactivation of the elastic response inside the contraction. 

On the other hand, when $\Delta p$ is reversed to a negative value, the flow lasts for a longer time and the fluid displaced through the contraction before shear jamming sets in again is about twice as much as that displaced in the first part of the experiment (Fig.~\ref{fig:ux}).
The shear-jammed domains, where the elastic response is active, are destroyed and rebuilt with a different spatial distribution by the reverse {flow (Fig.~\ref{fig:contraction}d).
The elastic stress at $t=\SI{25}{}$ {(Fig.~\ref{fig:contraction}e) sustains two subsequent pressure drops of about $\Delta p/2$, thus showing two pairs of equally stresses jammed domains.}
{At $t=\SI{8}{}$}, the total pressure drop is almost entirely sustained by the jammed domains on the left, while only a slight elastic response is visible on the right side of the contraction (Fig.~\ref{fig:contraction}c).}
By mimicking randomness in the suspension microstructure \citep{Nakanishi_2012} with spatial fluctuations of the initial deformation gradient, we obtained a more realistic nucleation of the shear-jammed regions.
{This random seed is at the origin of the asymmetry between the jamming process for positive or negative $\Delta p$.}
{Details on the dependence of simulation results on meshing and material parameters, together with movies of the time-evolution of pressure, flow, and elastic response are presented in the Supplemental Material}~\footnote{See Supplemental Material at [\emph{URL will be inserted by publisher}] for the derivation of Eq.~\eqref{eq:1d}, examples of identification of material parameters from experimental data, details on the dependence of simulation results on the meshing and on the values of the material parameters, and the full time-evolution of pressure, flow, and elastic response. It includes Refs.~\cite{Seto_2017,Seto_2018,Ness_2021}.}.


\medskip
\textit{Conclusions.}---%
We have shown how the knowledge about a complex collective phenomenon, acquired by means of experiments and simulations, can be transferred into a rather simple model of the macroscale physics that we observe.
By relating shear jamming to the activation of an elastic response and not to a divergence of the viscosity, we developed a tensorial model able to reproduce the qualitative features of shear jamming.

Such a model can be applied to generic flows and geometries in both two and three dimensions,
because it rests on physical considerations that are not peculiar to a specific experimental setup.
It becomes particularly useful to simulate the flow of suspensions in applications, where the focus is on the emergent collective physics and not on its microscopic origins. 

Notably, we simulated a material able to switch, in a reversible way, between a fluid-like and a solid-like behavior.
This feature, essential to capture shear jamming, can suggest effective ways to deal also with yielding phenomena.
While we kept the model as simple as possible, 
many  extensions can be implemented to reproduce a rate-dependent behavior.

\begin{acknowledgments}
G.G.G.\ acknowledges the support of the Italian National Group of Mathematical Physics (GNFM-INdAM) through the funding scheme \emph{GNFM Young Researchers' Projects 2020}.
R.S. acknowledges 
the support from the Wenzhou Institute, University of Chinese Academy of Sciences,
under Grants No.\,WIUCASQD2020002.
\end{acknowledgments}

%


\clearpage

\begin{widetext}

\begin{appendix}

\section*{\large Supplemental material for ``Shear jamming and fragility of suspensions in a continuum model with elastic constraints''}

\subsection{Equation for the strain in planar extension}

To derive Eq.~(6) in the main text, we start from the low-Reynolds-number approximation of the equation for unsteady flows with uniform gradient that reads
\begin{equation}
\rho\frac{\de \vc u}{\de t}=-\nabla p.
\end{equation}
Given that $\vc u(x,y,t)=(\edot(t) x,-\edot(t) y)$, we readily find $p(x,y,t)=\rho\ddot{\varepsilon}(t)(y^2-x^2)/2$.
The boundary condition at the outlets of the cross channel (Fig.~1, main text) imposes continuity of the traction, namely $\vt T\vc n$ (with $\vc n$ unit outer normal to the boundary) should balance the applied outward traction $\tau\vc n$.
If we now write that balance at $x=l$ and $y=0$, we obtain the only relevant stress component with the pressure term $\rho\ddot{\varepsilon} l^2/2$, the viscous term $2\eta D_{xx}=2\eta\edot$ and the elastic term 
\begin{equation}
2\kappa(\vt L-\Pi(\vt L))_{xx}=
\begin{cases}
  2\kappa [\varepsilon+r] & \text{if $\varepsilon <-r$},\\
                       0  & \text{if $-r \leq \varepsilon \leq r$}, \\
  2\kappa [\varepsilon-r] & \text{if $\varepsilon >r$}.
\end{cases}
\end{equation}
Hence, the balance equation becomes
\begin{equation}
\frac{\rho l^2}{2}\ddot{\varepsilon}+2\eta\dot{\varepsilon}-\tau
=
\begin{cases}
  -2\kappa [\varepsilon+r] & \text{if $\varepsilon <-r$},\\
                       +0  & \text{if $-r \leq \varepsilon \leq r$}, \\
  -2\kappa [\varepsilon-r] & \text{if $\varepsilon >r$}.
\end{cases}
\end{equation}

\subsection{Model parameters from particle-level simulations}

\begin{figure*}[b]
    \centering
    \includegraphics[width=8.6cm]{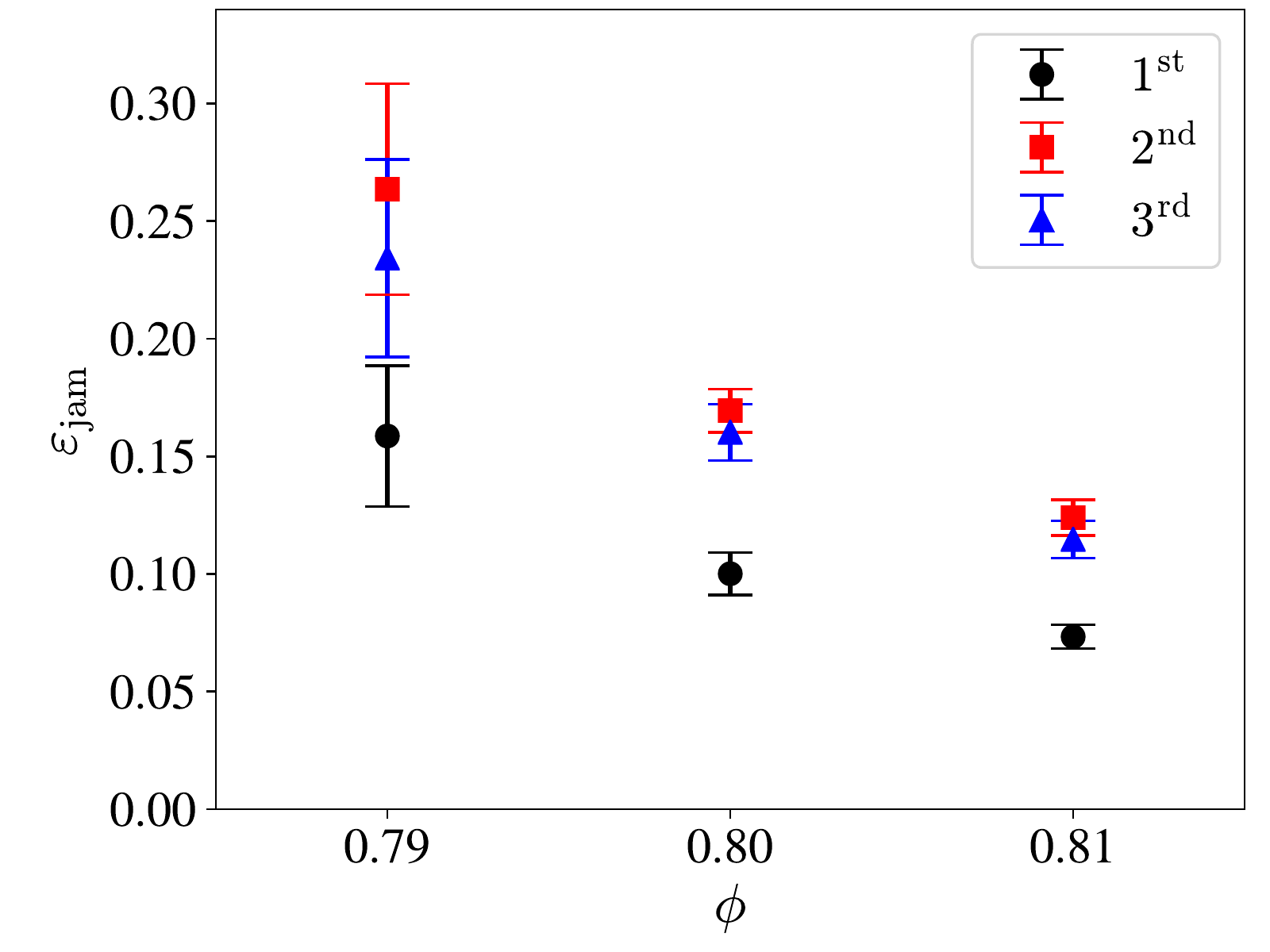}
    \includegraphics[width=8.6cm]{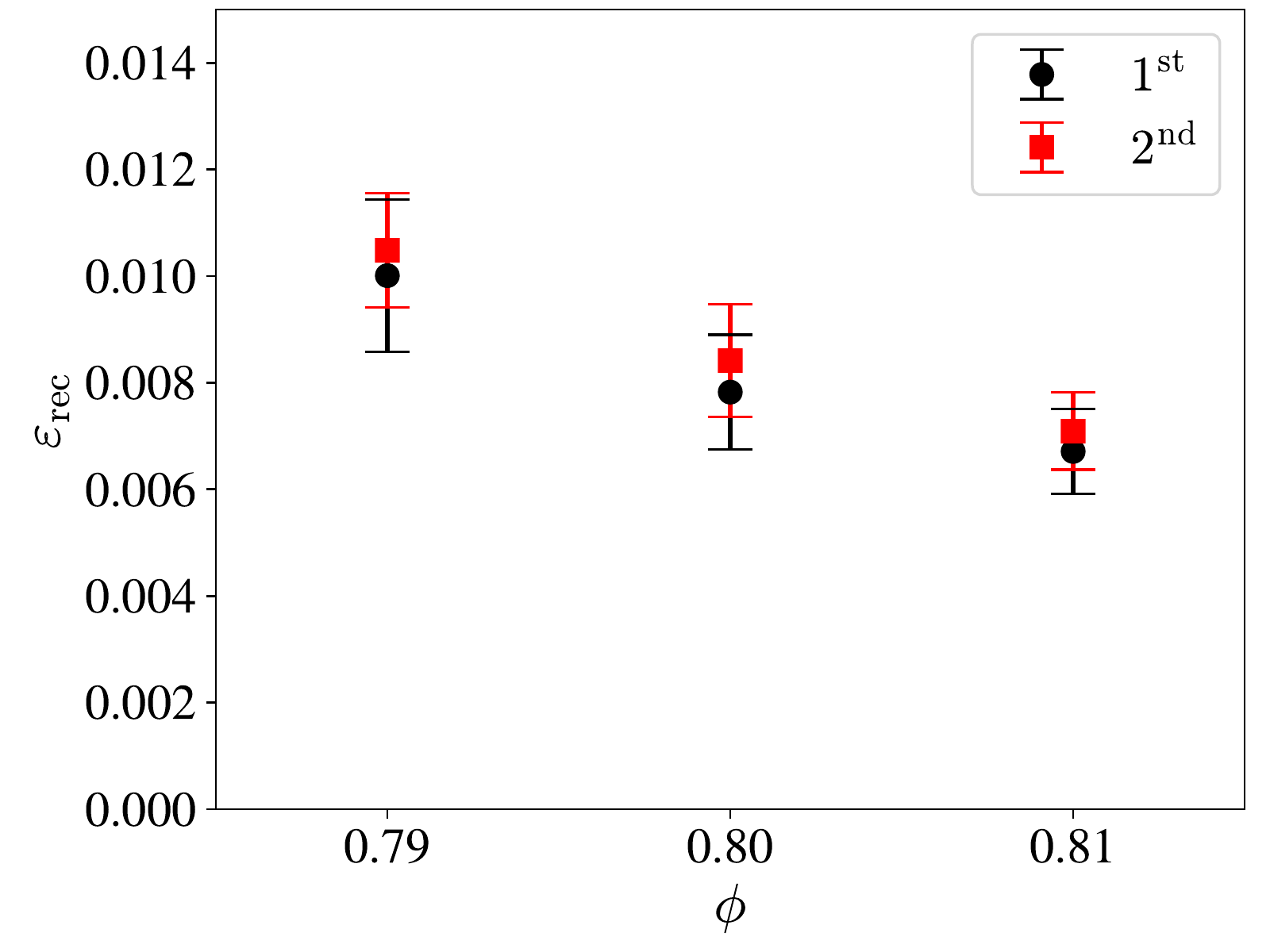}
    \caption{Data obtained from stress-controlled simulations of a two-dimensional monolayer system with three values of the area fraction $\phi$. The absolute value $\varepsilon_\mathrm{jam}$ of the strain accumulated between to different static configurations (left) is a decreasing function of $\phi$. Also the absolute value $\varepsilon_\mathrm{rec}$ of the strain recovered due to the elastic recoil upon removal of the applied stress is a decreasing function of $\phi$.}
    \label{fig:jamstrain}
\end{figure*}

Highly concentrated suspensions of solid particles dispersed in a viscous fluid
exhibit various non-Newtonian behaviors such as 
yield stress, shear thickening, and normal stress differences.
Recent particle simulations identified the basic microscopic contributions to these effects~\cite{Ness_2021}.
A common idealized model system 
is constituted by neutrally buoyant hard spheres suspended in a Newtonian liquid.
Hard spheres are rigid bodies and the excluded-volume interaction is purely geometric, without a characteristic force scale.
If inertia is negligible, 
the system has no specific force scale besides 
the one determined from an imposed flow.
The viscosity of such model suspensions is rate- or stress-independent.
If additional force scales are determined, for instance, by Brownian or repulsive interactions, the suspension rheology can be rate- or stress-dependent.

Following this hierarchy, to develop a constitutive model we conceptually start from the rate-independent case.
Such rate-independent particle simulations were used to investigate normal stress differences \cite{Seto_2018}
and shear jamming \cite{Seto_2019a}.
We here conduct particle-level simulations
for both simple shear and planar extensional flows~\cite{Seto_2017}.
The system is
 a two-dimensional monolayer system,
 containing small and large spherical particles 
 with the size ratio $1.4$.
 Areal fractions of the two populations are about half and half.
We set the friction coefficient $\mu = 1$
and the cutoff length scale 
$\delta/a = 10^{-2}$, with $a$ the radius of the small particles
(see Ref.~\cite{Mari_2014} for details about the computational method).

We performed stress-controlled simulations with three values of the area fraction $\phi$. We start from a well-relaxed initial state (prepared by a Brownian simulation)
and impose a given stress for a time sufficient to reach a first shear-jammed state. Then we remove the stress to observe the elastic recoil, and subsequently we reverse the initial stress to make the system flow again until another shear-jammed state is reached.
The protocol is repeated to observe two elastic recoils and three shear-jammed states.
Average values of the strain to jamming $\varepsilon_\mathrm{jam}$ and the recoil strain $\varepsilon_\mathrm{rec}$ are taken over 50 independent simulations.
The results presented in Fig.~\ref{fig:jamstrain} show that both $\varepsilon_\mathrm{jam}$ and $\varepsilon_\mathrm{rec}$ decrease upon increasing the volume fraction, as expected.
Since the initial state is not a jammed state, it is reasonable that the strain necessary to reach the first jamming be smaller that that necessary to go from one jammed state to the other.
On the other hand, the elastic recoil observed when removing the applied stress from a shear-jammed state is practically the same for every event.

In Fig.~\ref{fig:jamstrain} we report the results for the extensional flow simulation. Those obtained in simple shear are almost identical.
From these data, we can estimate the value of the parameter $r$ as one half of the strain between successive shear jamming events and that of the parameter $\tilde{\kappa}$ from the recoil strain \emph{via} the balance equation $2\tilde{\kappa}\varepsilon_\mathrm{rec}=\bar{\tau}$, with $\bar{\tau}$ the applied unit stress.

We estimate the values in the following table.
\begin{center}
\begin{tabular}{c||c|c|}
$\phi$ & $r$ & $\tilde{\kappa}$ \\
\hline
$\quad 0.79 \quad$ & $\quad 0.125 \quad$ & $\quad 50 \quad$\\
$\quad 0.80 \quad$ & $\quad 0.083 \quad$ & $\quad 62 \quad$\\
$\quad 0.81 \quad$ & $\quad 0.059 \quad$ & $\quad 72 \quad$
\end{tabular}
\end{center}

\subsection{Mesh-dependence of the numerical results}

To assess the relevance of our numerical results, with particular reference to the shape of the jammed domains in the clogging of a contraction flow, we performed finite element simulations using three unstructured meshes with increasing resolution in the central region where the channel width is reduced (Fig.~\ref{fig:mesh}).
The random initial condition on the strain tensor is generated on the larger mesh and then interpolated on the finer ones, to keep consistency of the evolution.

We found that the shape of the jammed domains and also the intensity distribution of the elastic response is well captured already with the intermediate mesh refinement (Fig.~\ref{fig:lambdamesh}), giving us confidence about the mesh-independence of the observed features.

\begin{figure}
    \centering
    \includegraphics[width=8.6cm]{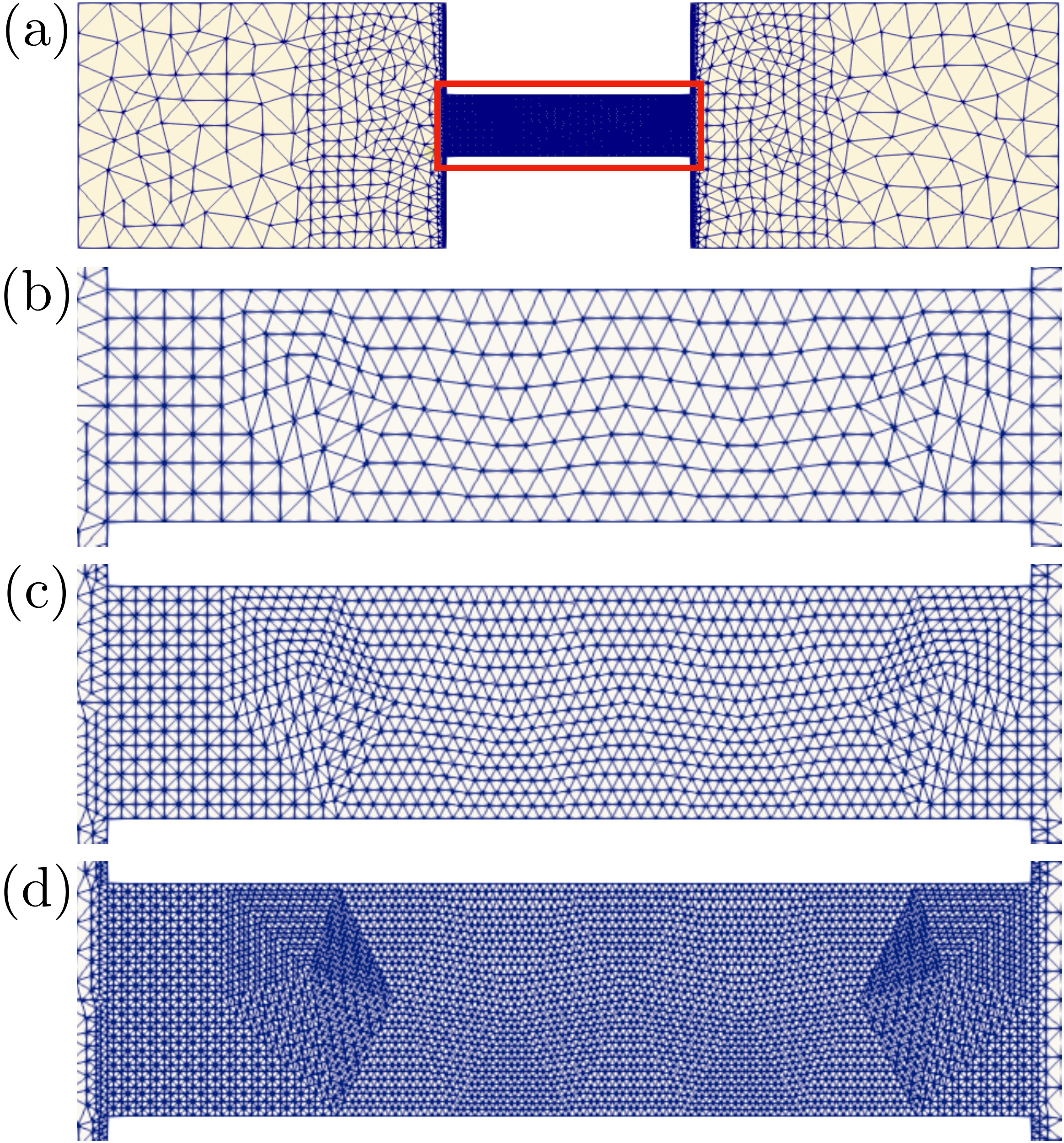}
    \caption{The entire domain considered for the contraction flow (a) is discretized with an unstructured triangular mesh. The mesh is not uniform to optimize the computational costs and is more refined in proximity of the contraction and within it. In the central region (red rectangle) we employed three different refinement levels (b--d) to assess the mesh-independence of the relevant features of our numerical solution of the flow equations.}
    \label{fig:mesh}
\end{figure}

\begin{figure*}
    \centering
    \includegraphics[width=18cm]{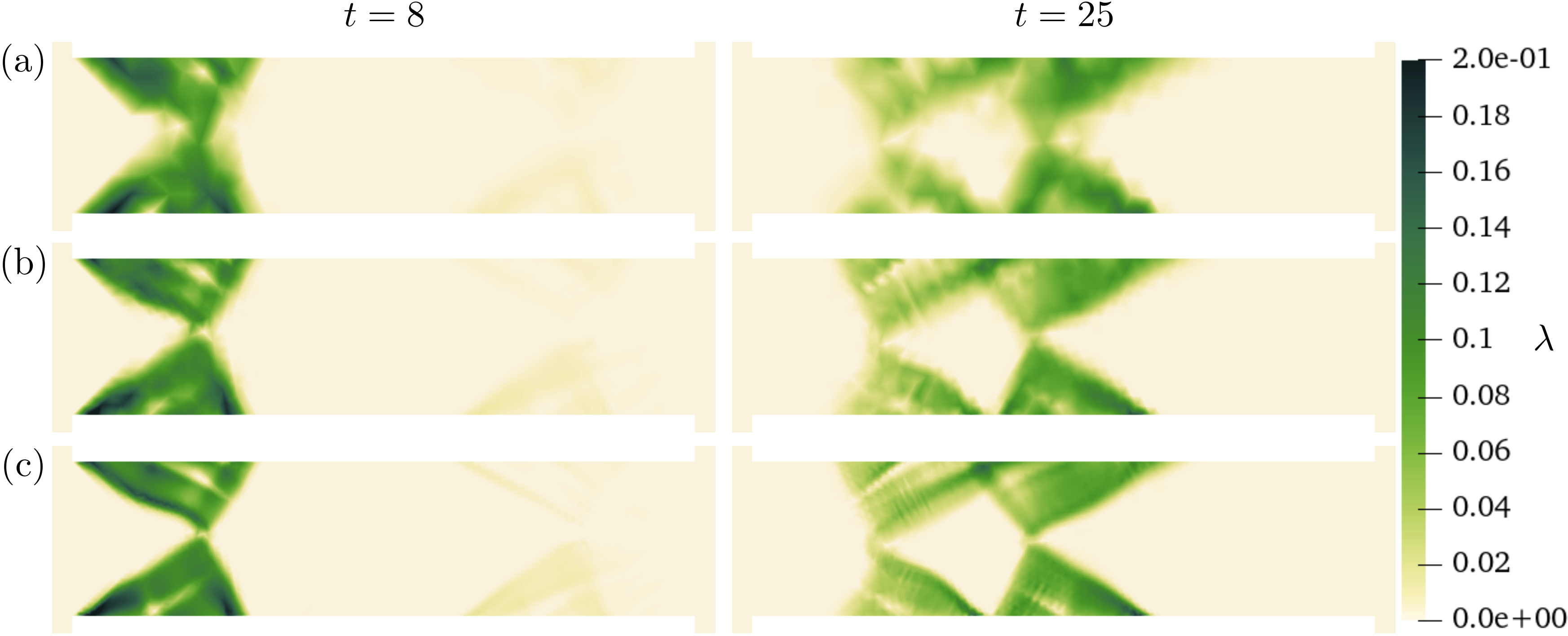}
    \caption{The distribution of the intensity $\lambda$ of the elastic response obtained in the jammed states of the clogging experiment described in the main text is compared in simulations with an increasingly refined mesh (from row (a) to (c)). With both positive ($t=8$) and negative ($t=25$) pressure difference we observe the stability under mesh refinements of the numerical results.}
    \label{fig:lambdamesh}
\end{figure*}

\subsection{Pressure field}

The time evolution of the pressure fields in the clogging--unclogging simulation can be viewed in the videos included in the Supplemental Material. 
During the flow, we observe a rather uniform pressure gradient that is distorted when the elastic response is activated. It is interesting to note that, in the clogged configurations, the jammed elastic walls sustain the pressure drops and separate basins with uniform pressure and vanishing velocity (Fig.~\ref{fig:pressure}). 

\begin{figure}
    \centering
    \includegraphics[width=8.6cm]{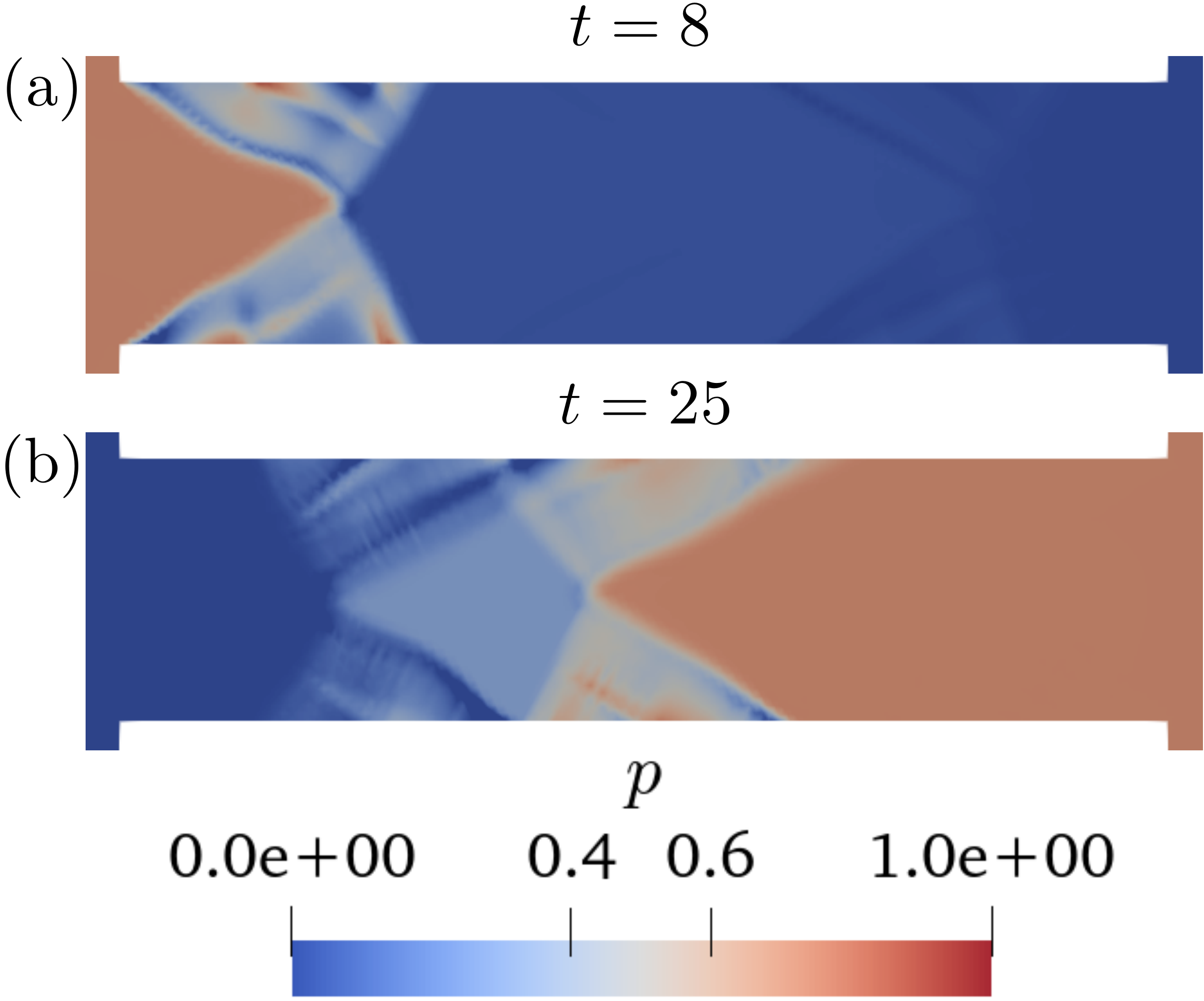}
    \caption{Considering the pressure distribution in the clogged states at $t=8$ and negative $t=25$, we see that the jammed elastic walls separate basins in which the velocity vanishes and the pressure is uniform. The elastic response sustains the pressure drops between the basins. Within the jammed domains we observe a fluctuating pressure distribution determined by the strain history of the incompressible elastic material produced by the jamming phenomenon.}
    \label{fig:pressure}
\end{figure}

\subsection{Dependence of the jamming dynamics on model parameters}

In this section, we briefly discuss how the jamming dynamics is changed by varying the radius $r$ of the elastically neutral set $\mathcal N$ and the elasticity constant $\tilde{\kappa}$. 

For the case of the extensional deformation presented in the main text, it is clear from Eq.~(6) that increasing $\tilde{\kappa}$ would decrease the amplitude of the elastic deformation at equal applied traction, producing a smaller recoil when the traction is removed. Moreover, it would increase the frequency of the elastic oscillations about the static equilibrium configuration in the transient motion. Decreasing $\tilde{\kappa}$ has the opposite effects.
We then explored the influence of $\tilde{\kappa}$ in the case of the contraction flow.
By varying $\tilde{\kappa}$ around the value ${\kappa}_0\approx 116$ (used in Figs.~2 and 3 of the main text) and performing simulations for $\tilde{\kappa}=\alpha{\kappa}_0$ for $\alpha\in\{0.1,1,10\}$, 
we found that, in spite of the higher complexity of the flow, the global effect on the macroscopic motion is completely analogous to the case of planar extension (Fig.~\ref{fig:kdependence}).

\begin{figure*}
    \centering
    \includegraphics[width=8.6cm]{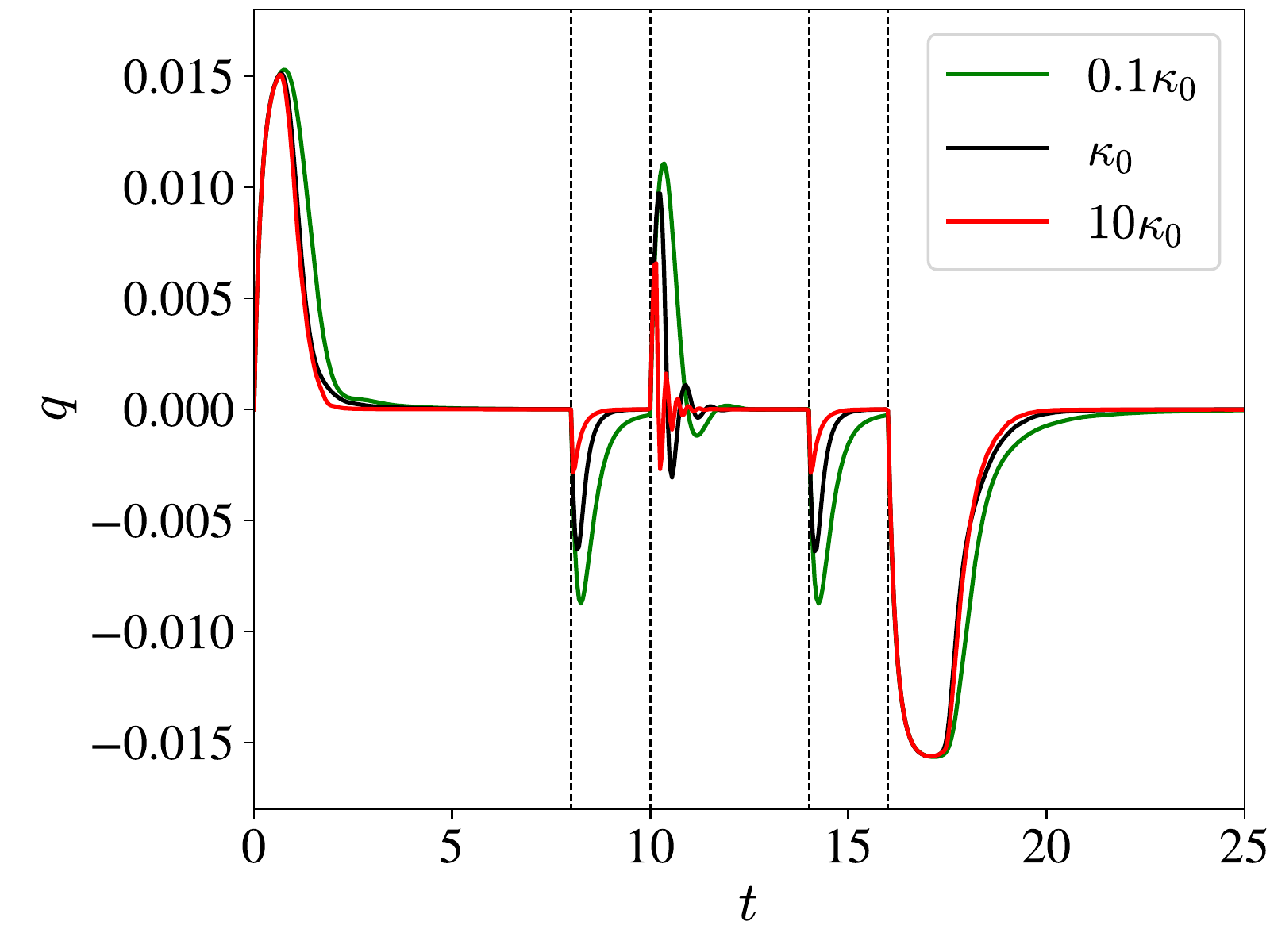}
    \includegraphics[width=8.6cm]{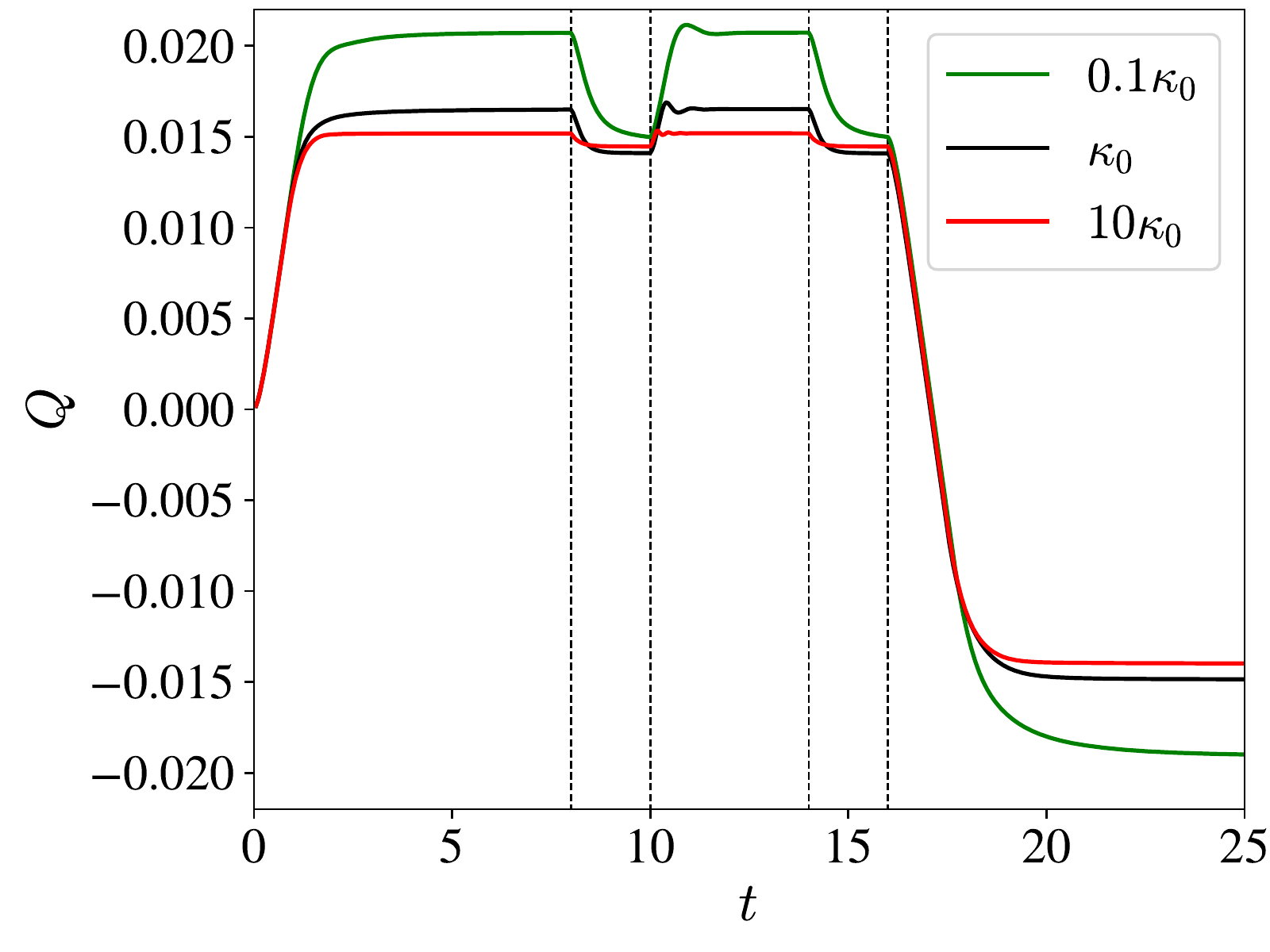}
    \caption{Time evolution of the flow rate $q$ through a section of the contracting channel and of the integrated quantity $Q(t)=\int_0^tq(s)ds$ computed by varying the elasticity constant $\tilde{\kappa}$ about the value $\kappa_0\approx 116$ used in Figs.~2 and 3 of the main text. We kept all of the other parameters fixed.
    By increasing (res.\ decreasing) $\tilde{\kappa}$ we decrease (resp.\ increase) the amount of elastic recoil (for $t\in[8,10]$ and $t\in[14,16]$) and the period of the oscillations about the static configuration observed immediately after $t=10$.}
    \label{fig:kdependence}
\end{figure*}

In very special uniform flows such as planar extension and simple shear, the dependence on $r$ is transparent. It measures how much we can deform the material before reaching jamming. 
Regarding the contraction flow, we explored the effect of varying $r$ and found that, on one hand,
if $r$ is small (namely comparable to the fluctuations in the initial microstructural  conditions) the jamming phenomenon becomes essentially dictated by this random initialization, because jamming takes place almost immediately, and the results loose their generic relevance.
If, on the other hand, $r$ grows larger, the interplay between the jamming threshold and the channel flow profile in the fluid regime leads to a modification of the shape of the jammed domains, with a more gentle slope on the high-pressure side of the elastically active domains (Fig.~\ref{fig:rdependence}).

\begin{figure*}
    \centering
    {\large
    $r=0.5/\sqrt{2}$
    \hspace{7cm}
    $r=3.0/\sqrt{2}$
    }
    \includegraphics[width=8.6cm]{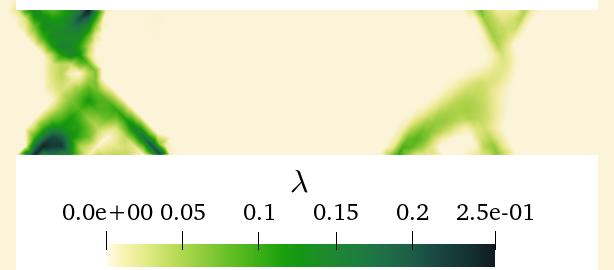}
    \includegraphics[width=8.6cm]{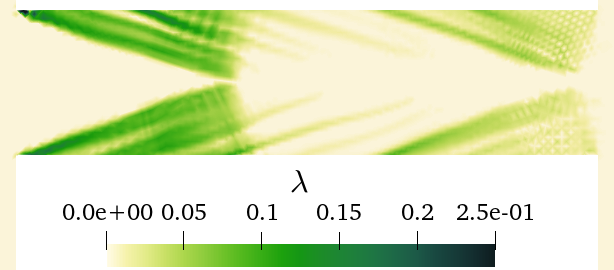}
    \caption{Intensity of the elastic response $\lambda$ measured in the jammed state at $t=8$ of the contraction flow computed for $r=r_0/3$ (left) and $r=2r_0$ (right), where $r_0=1.5/\sqrt{2}$ is the value used in Figs.~2 and 3 of the main text.
    We kept all of the other parameters fixed.
    For the smaller $r$, the shape of the jammed domains is rather asymmetric and essentially dictated by the random initial condition, because jamming occurs very soon in the simulation.
    For larger $r$, on the other hand, we can observe a modified sawtooth shape, similar to the one observed in Fig.~3, but with a more gentle slope on the high-pressure side.}
    \label{fig:rdependence}
\end{figure*}

\subsection{Videos}

The Supplemental Material includes two videos:
\begin{itemize}
    \item \texttt{Movie1.mp4} [\href{https://mediaspace.unipd.it/media/Shear+jamming+and+fragility+1/1_zzf8dmpe}{linked here}] presents the time-evolution of the pressure and velocity field in the contraction flow simulation from which Figs.~2 and 3 were extracted.
    \item \texttt{Movie2.mp4} [\href{https://mediaspace.unipd.it/media/Shear+jamming+and+fragility+2/1_oq4aqnyk}{linked here}] presents the time-evolution of the intensity of elastic response in the same flow.
\end{itemize}

\end{appendix}
\clearpage

\end{widetext}

\end{document}